\documentclass[twocolumn,preprintnumbers,amsmath,amsfonts,amssymb,notitlepage,showpacs,aps,prb,longbibliography]{revtex4-2}

\usepackage{color}

\date{\today}

\usepackage{amsmath}

\usepackage{graphicx}

\usepackage{hyperref}

\def \be{\begin{equation}}
\def \ee{\end{equation}}
\def \ba{\begin{array}}
\def \ea{\end{array}}
\def \bea{\begin{eqnarray}}
\def \eea{\end{eqnarray}}

\begin{document}
\title{
One-dimensional $Z_2$ lattice gauge theory
in periodic Gauss-law sectors
}
\author{Vaibhav Sharma}
\email{vs492@cornell.edu}
\author{Erich J Mueller}
\email{em256@cornell.edu}
\affiliation{Laboratory of Atomic and Solid State Physics, Cornell University, Ithaca, New York}

\begin{abstract}
We calculate the properties of a one-dimensional $Z_2$ lattice gauge theory in different {\em Gauss law sectors}, corresponding to different configurations of static charges, set by the orientations of the gauge spins.  Importantly, in quantum simulator experiments these sectors can be accessed without adding any additional physical particles or changing the Hamiltonian:  The Gauss law sectors are simply set by the initial conditions.  We study the interplay between conservation laws and interactions when the static charges are chosen to form periodic patterns.
We classify the different Gauss law sectors and use the density matrix renormalization group to calculate the ground state compressibility, density profiles, charge density wave order parameters, and single particle correlation functions as a function of matter density. We find confined and deconfined phases, charge density waves, correlated insulators, and supersolids. 
\end{abstract}

\maketitle

\section{Introduction}\label{intro}

Lattice gauge theories were developed to understand the models of elementary particle physics \cite{Zohar_2016}, and as a way to explore novel spin physics \cite{kogut}.  
Recently quantum simulator experiments have produced the first 
physical realizations of these models \cite{exp5,sccircuit,floquet}, and begun to experimentally study their properties.  
Here we theoretically explore the behaviour of one of these novel systems, the realization of a
1D $Z_2$ lattice gauge theory.  
This model
has an extensive set of conserved quantities, dividing the Hilbert space into disconnected and non-equivalent sectors.  Previous theoretical work has mostly focussed on the low density regime of only one of these sectors, referred to as the uniform configuration.  
Here we present a global view, describing the properties of the system as a function of matter density for different sectors, corresponding to adding a periodic array of immobile charges.  We find that the model is much richer in these non-uniform configurations, displaying commensurate and incommensurate correlated insulators, dimerized fluids and supersolidity.  The 
sector can be set by initial conditions in an experiment and hence one can access all of this rich physics without changing the Hamiltonian and maintaining its translational invariance.  

One important application of physical realizations of lattice gauge theories is 
as {\em quantum simulators} which address  questions about models of interest to fundamental physics.  They could be used in a way that is similar to classical numerical simulations but with different strengths.  For example, real-time dynamics are 
challenging to explore with numerical simulations but are 
natural in cold atom experiments.   
Furthermore, simulating these theories with quantum hardware could help us uncover new phenomena and probe regimes which are currently inaccessible.
We emphasize, however, that these experiments are valuable apart from their connection to high energy physics:  
They display a rich set of many-body phenomena, 
including the ones studied here.


As typically formulated \cite{Zohar_2016}, a lattice gauge theory has matter degrees of freedom on a lattice and  gauge degrees of freedom  on the bonds.  The gauge fields define the rules for parallel transport of the matter fields.  The model is invariant under rotating the local Hilbert space of any given site and the surounding bonds.  These local gauge transformations form a group.  In lattice versions of Quantum Electrodynamics (QED) and Quantum Chromodynamics (QCD), these groups  are  continuous (isomorphic to $U(1)$ and $SU(3)$ respectively).  There are also lattice gauge theories with discrete symmetry groups.  Here we focus on an experimentally relevant and simple case where the gauge group is $Z_2$.


In realizing their $Z_2$ lattice gauge theories, the experiments partially fix the gauge.  They effectively work in the {\em temporal gauge}, where the time-like component of the gauge fields vanish.  In the temporal gauge, Gauss' law is enforced as a conservation law:  If it is satisfied at time $t=0$, it will be satisfied for all time.  Different initial conditions can therefore lead to different variants of Gauss' law, and we refer to these as Gauss law sectors.  As already explained, they differ by the presence of static charges.  The sector with no static charges will be referred to as the uniform sector.

One dimensional $Z_2$ lattice gauge theories have been well studied within the uniform sector \cite{atomsmeetlgt,z2dimers}.
Those
studies found that at low densities the 
matter degrees of freedom are confined, forming superfluid bosonic dimers~\cite{z2dimers}.  Adding longer-range matter interaction terms can lead to other phases such as charge density waves or supersolids~\cite{z2interactions,z2confined,devilstaircase}.  One of our main results is that the additional long-range interactions are unneccessary for achieving these exotic phases.

Non-uniform Gauss law sectors have received very little attention.  The one notable exception is the concept of ``disorder-free localization'' where
%
%
initial states prepared in a superposition of different Gauss law sectors 
experience
an effective binary disorder potential. This randomness localizes the matter particles, despite the fact that the Hamiltonian is translationally invariant  
\cite{dl2,dl,dl1,dl3,dl4,dl5}. 

On the experimental side there has been substantial progress in the realization of $Z_2$ and $U(1)$ lattice gauge theory models. Superconducting quantum circuits have been used to observe gauge invariance and confinement dynamics in a $Z_2$ lattice gauge theory~\cite{exp2,sccircuit,exp4}. Cold atom based quantum simulators have implemented $Z_2$~\cite{floquet} and $U(1)$~\cite{exp5} lattice gauge theories, and probed the thermalization dynamics of a $U(1)$ gauge theory~\cite{exp6}.



\begin{figure}
\includegraphics[width=\columnwidth]{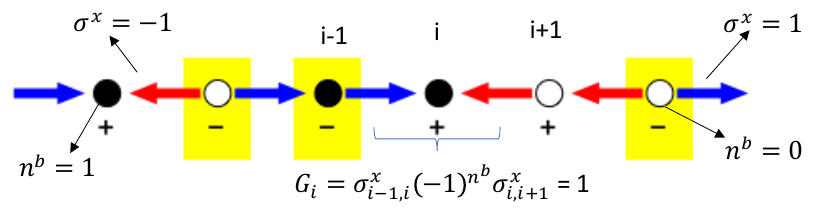}
    \caption{A basis state for the 1D $Z_2$ lattice gauge theory model. 
     Sites are designated by either filled or empty circles, representing the presence or absence of a boson.  The link between site $i$ and $i+1$ contains a spin-1/2, whose direction, $\sigma^{x}_{i,i+1}$ is designated by either a right-pointing blue arrow or a left-pointing red arrow.
     Each site $i$ has a conserved quantity, $G_i = \sigma^{x}_{i-1,i}(-1)^{n^b_i}\sigma^{x}_{i,i+1}$. Sites where the conserved quantity $G_i = -1$ have a yellow colored box around them.}
    \label{lattice}
\end{figure}

\section{Model}\label{model}

Our model contains hard-core bosons hopping between the sites of a 1D lattice, interacting with spin-1/2 degrees of freedom that sit on the bonds.  Figure~\ref{lattice} shows a typical state in the Hilbert space:  open or filled circles correspond to empty or full sites, while left-pointing red arrows and right-pointing blue arrows represent the two basis states of the gauge spins.  The Hamiltonian is,
\begin{equation}\label{hamiltonian}
    {H} = -t \sum_{i} (b_{i+1}^{\dag}\sigma^{z}_{i,i+1} b_{i} + h.c.) -h\sum_{i} \sigma^{x}_{i,i+1}.
\end{equation}
Here $b,b^{\dagger}$ are annihilation and creation operators of hardcore bosons on the sites, $i$ is the site index and $\sigma^x,\sigma^z$ are Pauli operators for the spin-1/2 particles on links. The lattice spacing $a$ is set to unity. Throughout this work, we set $t=1$. When a boson hops, it flips the spin on the link.  This is the $Z_2$ analog of minimal coupling where one interprets $\sigma_z 
=e^{i \pi/2 (\sigma_z -1)}
=e^{i A_1}$, where $A_1$ is the spatial component of the vector potential.  In this interpretation the electric field, which is conjugate to the vector potential, is  $E=(\sigma_x+1)/2$.  The electric field takes on the values $0,1$, and the Zeeman term can be  interpreted as the energy associated with this electric field.  In this formulation there is no time-like component of the vector potential, as one assumes the temporal gauge $A_0=0$.



Our Hamiltonian is invariant under the group of local spatial gauge transformations, generated by $G_i = \sigma^{x}_{i-1,i}(-1)^{n^b_i}\sigma^{x}_{i,i+1}$, where $n^b_i = b_{i}^{\dag} b_i$ denotes the number of bosons on site $i$.  The various operators in the Hamiltonian transform as $G_i b_i G_i^{\dagger} = e^{i \pi} b_i$, $G_i b_i^{\dagger} G_i^{\dagger} = e^{-i \pi} b_i^{\dagger}$ and $G_i \sigma^z_{i,i+1} G_i^{\dagger} = e^{i \pi} \sigma^z_{i,i+1} $.   These are the $Z_2$ analogs of $b_i\to e^{i\alpha_i} b_i$ and $A_{ij}\to A_{ij}+\partial \alpha_i-\partial \alpha_j$.
It is straightforward to show that $[G_i,H]=0$.




The eigenstates of $H$ can be chosen such that they are also simultaneously the eigenstates of each of the $G_i$, so that $G_i |\psi\rangle = e^{i\pi \eta_i} |\psi\rangle$ where $\eta_i=0,1$.  This can be interpreted as a form of Gauss' law,  $\sigma_{i-1,i}^x \sigma_{i,i+x}^x |\psi\rangle= e^{i\pi/2 (\sigma_{i-1,i}^x -\sigma_{i,i+x}^x)} |\psi\rangle = G_i (-1)^{n^b_i}|\psi\rangle = e^{i\pi(n^b_i+\eta_i)} |\psi\rangle$, which relates the difference of the ``electric field'' on two consecutive links to the effective charge on site $i$, $E_{i.i+1}-E_{i-1,i}=q_i=n^b_i+\eta_i$ (mod 2).  This restriction to a space of fixed $\eta_i$ is referred to as choosing a Gauss law sector, and corresponds to selecting the locations of static background charges.
We will use a list of $+$ and $-$ signs (corresponding to $\eta_i = 0$ and $\eta_i = 1$ respectively) to designate the signs of sequential $G$'s.  We emphasize that these sectors are physically distinct:  As we describe below, different choices for $\{G_i\}$ yield different behavior.
Most previous works have studied the {\em uniform sector} where $G_i = 1$ ($\eta_i =0$) for all $i$.  In an experiment, the sector is chosen by initial conditions of the gauge spins and no additional physical particles need to be added.

One useful observation is that the model in Eq.~(\ref{hamiltonian}) is invariant under a particle-hole transformation where $b_i \to b^{\dag}_i$ and $b^{\dag}_i \to b_i$. The boson number density maps to hole density as, $n_i^b \to (1-n^b_i)$. For hardcore bosons, $(-1)^{n_i^b} = (1-2n^b_i)$ and thus the conserved quantities, $G_i \to -G_i$ after the particle-hole transformation.




When $h=0$, the model in Eq.~(\ref{hamiltonian}) 
can be mapped onto a hopping-only model of hard core bosons by writing
$a_i = b_i\prod_{j\geq i}\sigma^z_{j,j+1} 
$ and $a_i^{\dag} = b_i^{\dag}\prod_{j\geq i}\sigma^z_{j,j+1}
$. 
One then finds for any $\{G_i\}$, $H=-\sum_i
(a_{i+1}^\dagger a_i + a_i^\dagger a_{i+1})$.  This can further be mapped onto free fermions via a Jordan Wigner transformation.  

The Zeeman term will introduce long range interactions between these fermions.  The structure of this interaction term depends on which sector one is in.  In particular, consider an open chain where the first gauge spin is aligned in the positive $x$ direction, $\sigma_{0,1}^x=+1$.  Gauss' law then uniquely constrains 
$\sigma^x_{i,i+1}=\prod_{j\leq i} (-1)^{n_j} G_j$.  This can be substituted into $H$ to give a model that only depends on the transformed bosons ($a_i$).  The Zeeman term becomes a non-local many-body interaction of the form
\begin{equation}\label{interaction}
V=-h\sum_{i}\prod_{j\leq i}(-1)^{n_j} G_j. 
\end{equation}
We will take the $G_j$ to form a periodic pattern.
The interplay between this periodicity and
strong interactions leads to novel physics. In Appendix~\ref{isss} we show that for periodic patterns of $G_j$'s with ``inversion-shift" symmetry, this term can be further expressed as just a two-body finite range hard-core interaction term added to a periodic lattice potential.




\begin{figure*}
    \includegraphics[width = 18cm,height=8.5cm]{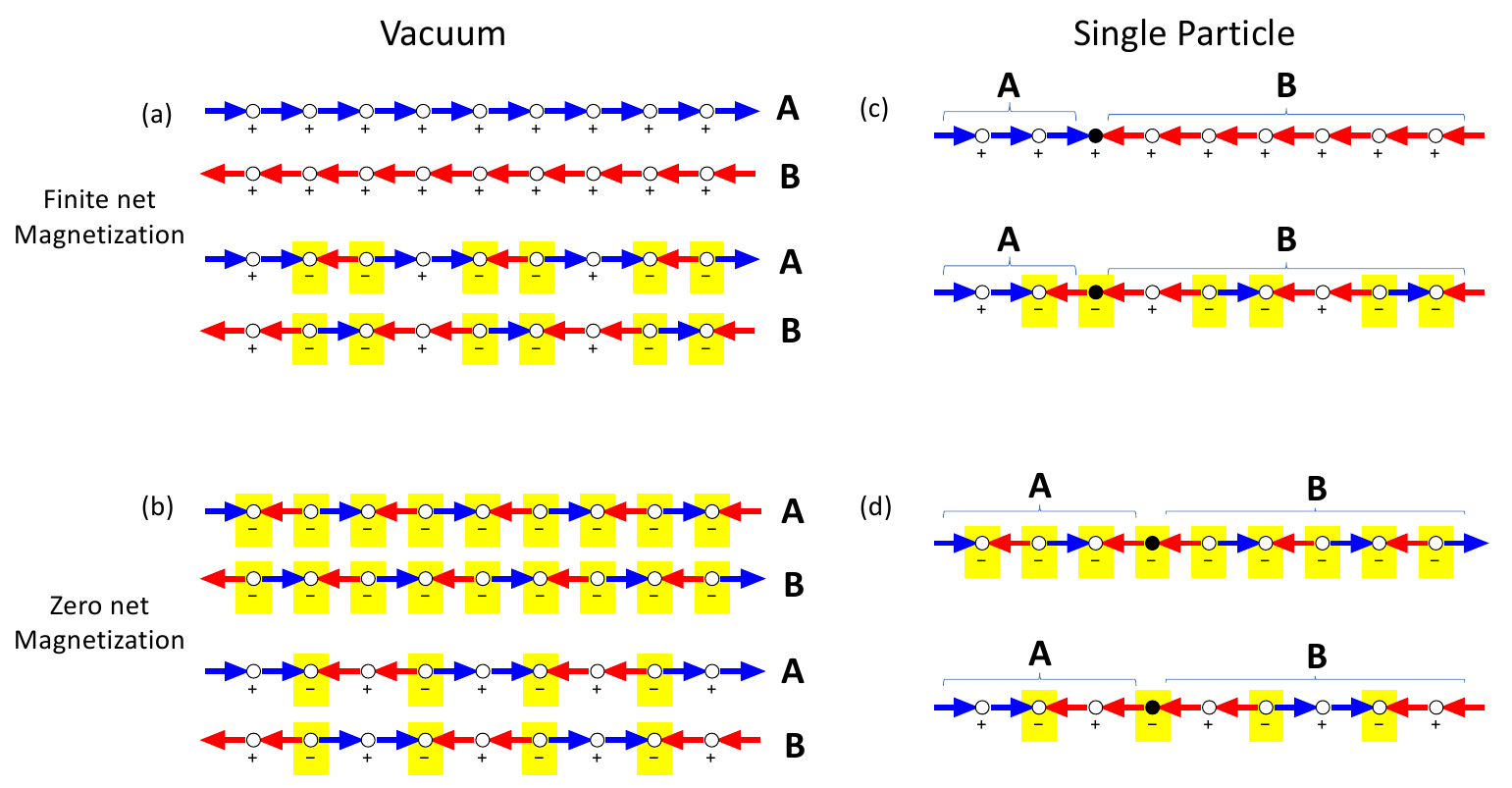}
    \caption{(a),(b) Gauss laws sectors with (a) finite  and (b) zero  net magnetization 
    in the absence of any bosons (vacuum). The A and B phases correspond to the two parity reversed vacuum states.  In the main text these configurations would be written as $G=+++++,+--+--,-----,+-+-+-$. (c) A single boson in a finite magnetization sector forms a domain wall between the two parity reversed spin configurations. Shifting the boson costs finite energy. (d) In a zero magnetization sector, the boson and the domain wall can be shifted by a unit cell without an energy cost.}
    \label{gaugeplot}
\end{figure*}

\section{Results}\label{sec:results}

Our primary goal is to understand the properties of Eq.~(\ref{hamiltonian}) for different choices of the conserved quantities $\{G_i\}$.  We will restrict ourselves to periodic arrangements with small unit cells. 
We will denote these by showing a few unit cells, such as $+++++$ or $+-+-+-$.  An important characteristic is the fraction of the $G_i$'s which are $-1$, a quantity we denote $n_m$. 

In the absence of particles, the eigenstates of $H$ consist of patterns where each spin has a definite $x$-orientation.  Up to a global spin flip, the pattern is fully defined by the $G_i$'s.
When $G_i=1$ the two spins on either side of the site will be aligned, while when $G_i=-1$ they are anti-aligned.
The  energy of each state is simply proportional to its net magnetization.
Arrangements where the unit cell contains an
odd number of $-$'s will always have zero net magnetization, and hence zero energy.  This is illustrated in Fig.~\ref{gaugeplot}(b).  We will 
consider patterns with relatively simple unit cells, 
in which case having an even number of $-$'s results in a non-zero magnetization.  The case of the uniform configuration, where the unit cell contains zero $-$'s is an example. 
Figure~\ref{gaugeplot}(a),(b) shows both vacuum spin configurations which are consistent with each Gauss law choice.  When the net magnetization is non-zero, these have different energies.   

We can now consider adding a single boson.
Due to Gauss' law, that boson acts as a domain wall, separating two parity reversed vacuum states. This is shown in Fig.~\ref{gaugeplot}(c),(d).  If the vacuum has a net magnetization, then the particle will feel a force proportional to $h$:  It lives in a tilted potential landscape.  Therefore the spectrum will consist of a Wannier Stark ladder, and in the time domain one will observe Bloch oscillations \cite{wannierstark}.  The particle is localized. Conversely, if the vacuum has no net magnetization, then the particle simply experiences a periodic potential, and the spectrum contains extended Bloch states. 

Writing the two vacuum states as $A$ and $B$, the spin state with two bosons will have the form $ABA$, with two domain walls.
When the vacuum magnetization is finite, 
the energy density of the $B$ region is larger than the $A$ region, binding pairs together.
This phenomenon has been well described in the uniform sector \cite{z2dimers}.
The energy of a pair is linear in the separation between the bosons.  When the vacuum magnetization vanishes, this pairing mechanism is absent.  

This pairing is analogous to Quark confinement in QCD, and these paired phases are referred to as {\em confined}.  A simple example is contrasting the $+++++$ and $-----$ sectors.  In the absence of particles the former has a net magnetization, while the latter does not.  Thus for infinitesimally low densities one expects them to respectively be confined and deconfined.  These sectors are, however, connected by a particle-hole transformation.  Thus 
as the density approaches $1$
we expect a deconfined phase in the $+++++$ sector. 

One particular caution is that these simple arguments only apply to the case of a vanishingly small density of particles or holes.  The properties at finite densities are not obvious.  In all of the sectors that we have explored, we find that the particles are always confined at finite particle/hole density.


We distinguish between the confined and deconfined phases by studying the properties of the single particle density matrix $C_{ij}=\langle a_i^\dagger a_j\rangle$.  In the confined phase this density matrix should fall off exponentially with a finite correlation length, while in the deconfined phase one expects power law correlations with a diverging correlation length. To calculate the density matrix, we use an infinite matrix product state ansatz for the ground state wavefunction \cite{idmrg}.  In Sec.~\ref{sec:correlations} we use the VUMPS algorithm \cite{vumps} using the ITensor library to optimize this wavefunction and then calculate the correlation functions.



\begin{figure}
    \includegraphics[width=\columnwidth]{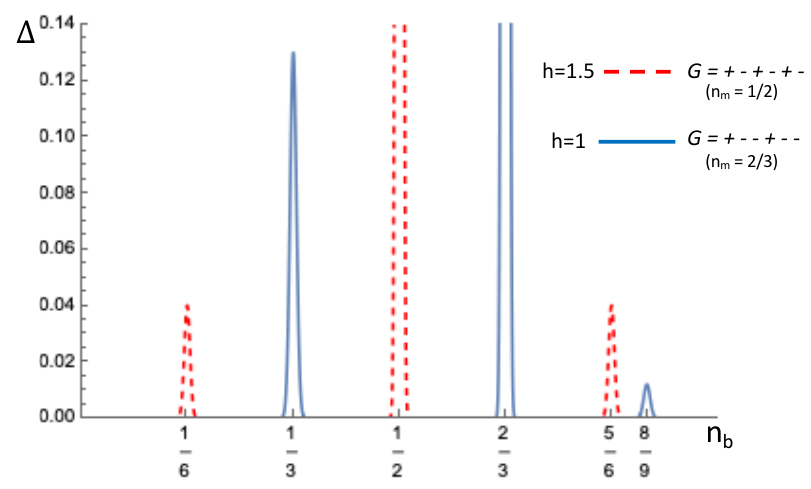}
    \caption{The dimensionless charge gap, $\Delta=(E(N+2)+E(N-2)-2 E_N)/2t$ (with $t=1$) extrapolated to $1/L\to 0$ at various rational boson fillings, $n_b$ for the Gauss law patterns, $G^{(1)} = +-+-+-$ (red dashed) and $G^{(2)} = +--+--$  (blue solid) at $h=1.5$ and $h=1$ respectively. The largest gaps are off-scale:  $\Delta\approx4h,6h$ at $n_b=1/2,2/3$ for $G^{(1)}$ and $G^{(2)}$.}
    \label{fig:gap}
\end{figure}

Before giving our results for the single particle correlations we first argue for the presence of incompressible states at particular densities.
We numerically search for these insulating states by calculating the ground state energy of $N$ particles in a system with $L$ sites, $E_L(N)$.  
The chemical potential in the thermodynamic limit is $\mu=E_L(N+1)-E_L(N)$.  An incompressible (insulating) state is characterized by a discontinuity in the chemical potential, 
characterized by the dimensionless gap
$\Delta=\lim_{L\to \infty} (E_L(N+2)+E_L(N-2)-2 E_N)/2t$. We use the finite Density Matrix Renormalization Group (DMRG) to calculate $E_L(N)$ for several different values of $L$, and then extrapolate to the thermodynamic limit by taking the intercept as $1/L \to 0$. Figure~\ref{fig:gap} shows representative examples, which will be discussed in detail below.
 As is known from previous work \cite{z2dimers}, in the uniform sector $+++++$ or its particle-hole transform, $-----$, the only insulators are the trivial ones where there is zero or one particle per site.

 In Fig.~\ref{fig:gap} we see gapped states at various rational values of $n_b$ where $n_b = N/L$.  The largest gaps are seen when the density of particles is equal to $n_m$, the fraction of sites with $G_i=-1$.  We refer to the ratio $\nu=n_b/n_m$ as the filling.  In Sec.~\ref{sec:peierls} we discuss the integer filling insulators (commensurate insulators), where $\nu=1$.  In Sec.~\ref{secother} we discuss the fractional filling insulators (incommensurate insulators), where $\nu$ is a rational number with a non-trivial denominator.

 \begin{figure}
    \includegraphics[width=\columnwidth]{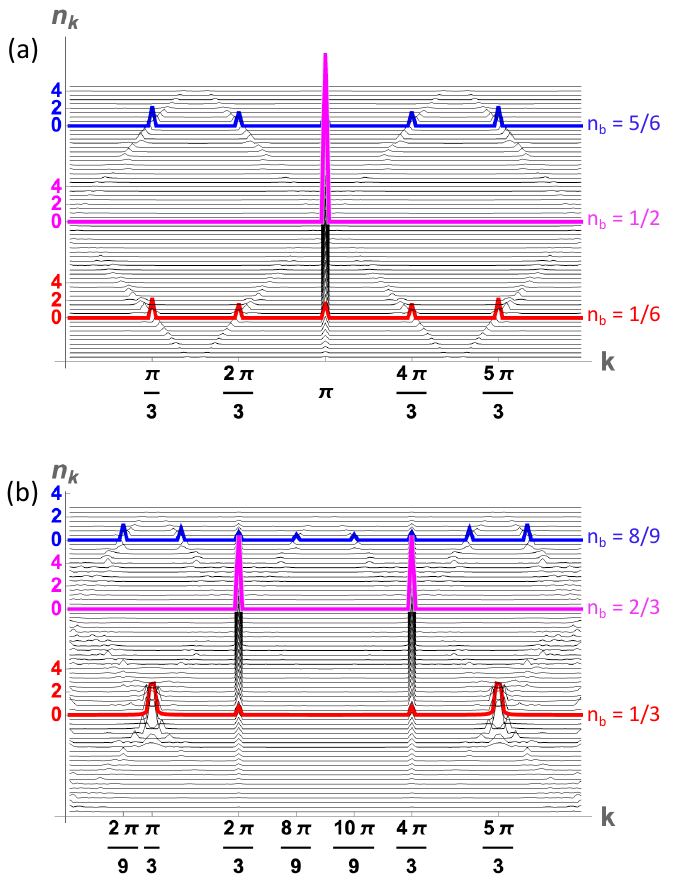}
    \caption{(a),(b) contain multiple curves (vertically separated by arbitrary amount) showing the fourier transform of spatial charge density in the ground state at various boson fillings in the Gauss law pattern, $G = +-+-+-$ (zero net magnetization) and $G = +--+--$ (finite net magnetization) respectively. Colored highlighted curves show fourier spectra corresponding to charge density wave ordering at $n_b = 1/6,1/2,5/6$ in (a) and at $n_b = 1/3,2/3,8/9$ in (b).}
    \label{fig:fourier}
\end{figure}

These insulating states correspond to density waves.  Thus they can also be characterized by the Fourier transform of the density $n_k$.  Figure~\ref{fig:fourier} shows $n_k$ for various fillings in the $+-+-+-$ and $+--+--$ sectors.  The curves for the incompressible states, displayed in color, show significant large peaks at the corresponding wave vectors.  

\subsection{Integer filling insulator states}\label{sec:peierls}

\subsubsection{Instability towards a commensurate charge density wave}

As seen in Fig.~\ref{fig:gap} we find an insulator when the density of bosons $n_b$ is equal to the density of  $G_i=-1$ sites, $n_m$; $\nu=n_b/n_m=1$.  For example, when $G=+-+-+-$ there is a large charge gap when $n_b=1/2$, while when $+--+--$ there is a gap at $n_b=2/3$.  For moderate or large $h$, as shown in the figure, these gaps are of order $h$.  The figure also shows smaller gaps, corresponding to insulators which will be discussed in Sec.~\ref{secother}.  We identify the transition into the commensurate insulators as a Peierls instability, and below we argue that this transition occcurs at infinitesimal values of $h$.
This is true for patterns with both zero and finite net magnetization.

\begin{figure}
    \includegraphics[width=\columnwidth]{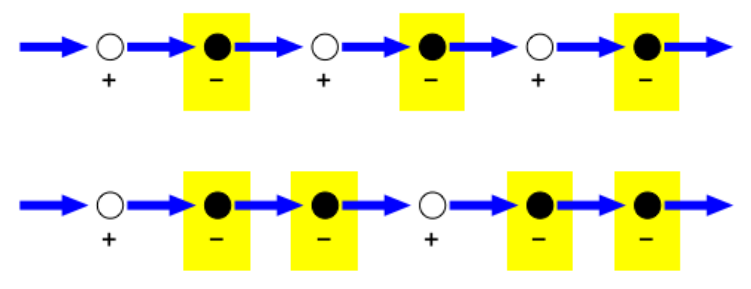}
    \caption{Ground state with all gauge spins aligned when bosons occupy '-' sites in the, (a) zero net magnetization sector and  (b) finite net magnetization sector}
    \label{peierls}
\end{figure}

The structure of these charge-density waves is best illustrated by the example in the top figure of Fig.~\ref{peierls} where 
$G_i = (-1)^{i}$.  If we place a boson on each of the $G_i=-1$ sites, then all of the spins become aligned.  For large $h$ this is a low energy state. Fig.~\ref{peierls}(b) gives another example, and it should be clear that in any sector, the Zeeman energy is minimized by placing the bosons on the $G_i=-1$ sites.


\begin{figure}
    \includegraphics[width=\columnwidth]{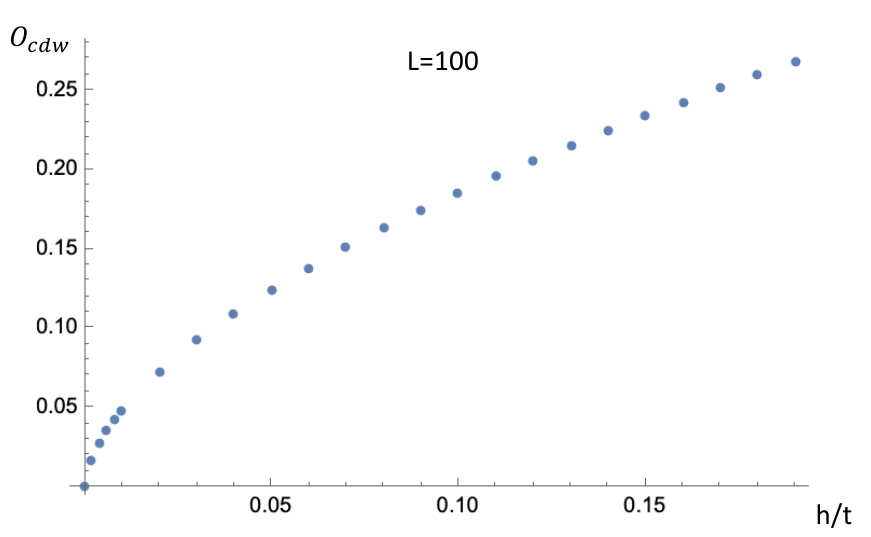}
    \caption{Magnitude of the charge density wave order parameter, $O_{cdw} = 2/L \sum_i (-1)^{i}n^b_i$ as a function of field $h/t$ in the Gauss law pattern, $G = +-+-+-$ using DMRG for a system size $L=100$ sites at half filling. As $h/t \to 0$, the order parameter's slope diverges.}
    \label{ocdw}
\end{figure}

We use bosonization arguments to analytically show that the Peierls transition happens at infinitesimal $h$.  For simplicity we take $G_i=(-1)^{i}$ but the argument clearly generalizes.  We eliminate the gauge spins by using the transformation in Sec.~\ref{model}, and note that in this sector $\Lambda_i=\prod_{j\leq i} G_j=\sqrt{2}\cos( \pi (2i-1)/4)$.  We then write the $V$ in Eq.~(\ref{interaction}) as $V=-h\sum_i \Lambda_i \prod_{j\leq i} (-1)^{n_j}$.  Noting that for integer $x$, one has $(-1)^x=\cos \pi x$, we then find
\begin{multline}\label{nofield}
    {H} = -\sum_{i} (a_{i+1}^{\dag} a_{i} + h.c.) \\
    -\sqrt{2}h\sum_{i} \cos{\frac{\pi (2i-1)}{4}} \cos\left(\pi{\sum_{j \leq i} n^a_j}\right).    
\end{multline}
We now go to the continuum limit, defining
$n_j = \rho(x) = \rho^0 - \partial_x \phi/\pi$.   At half-filling, $\rho^0=1/2$, the Zeeman term in the Hamiltonian in Eq.~\ref{nofield} becomes,
\begin{eqnarray}
   H_h
   &=&-\sqrt{2}h \int dx  \cos \frac{\pi(2x-1)}{4} \cos {\left(\frac{\pi x}{2}-\phi(x)\right)}.
\end{eqnarray}
Expanding the trigonometric functions we find a RG relevant term \cite{Giamarchi} of the form
$H_{\rm CDW}\propto h \int dx \cos{\phi(x)}$. Thus an arbitrarily small $h$ leads to a
commensurate charge density wave with a gapped spectrum. We numerically confirm this by using DMRG to calculate the ground state of a 100 site system. In Fig.~\ref{ocdw}, we show the commensurate charge density wave order parameter, $(O_{cdw} = 2/L \sum_i (-1)^{i}n^b_i)$ as a function of field $h$. We see that the order parameter has an infinite slope as $h \to 0$ and thus the susceptibility, ${\partial O_{cdw}}/{\partial h}$ diverges when $h \to 0$. The same structure is found in other sectors for $\nu=1$.

\subsubsection{Excitations}\label{ex}

As with particles in the uniform sector, the charged excitations of these Peierls insulator states live in a ``tilted'' energy landscape.  For example, removing a single particle from the states in Fig.~\ref{peierls} results in a spin configuration with a domain wall:  To the left of the missing particle the spins are aligned in the $+x$ direction, while to the right they are aligned in the $-x$ direction.  This costs an extensive energy, and the hole feels a force, moving it to the right.  Adding a particle has the same effect.  

\begin{figure}
    \includegraphics[width=\columnwidth]{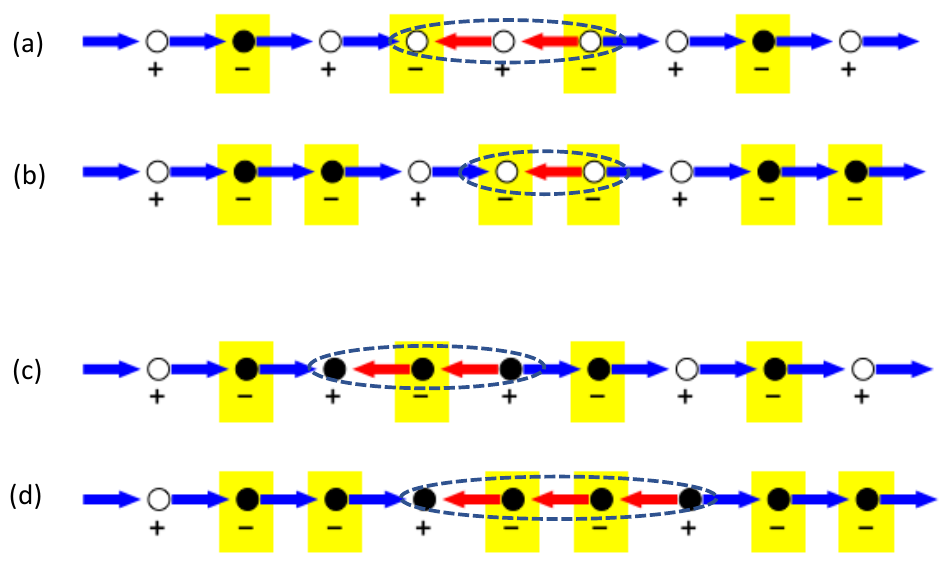}
    \caption{ Configurations from Fig.~\ref{peierls} (showing integer filling Peierls insulators) with (a,b) two holes or (c,d) two particles added.
    A string of flipped spins extend between the added particles/holes.}
    \label{pairing}
\end{figure}

One consequence is that particle or hole excitations of the Peierls insulators will pair up.  This is illustrated in Fig.~\ref{pairing}.  The energy of two holes (or particles) is linear in their separation, and the energy is minimized when they are as close together as possible.  This corresponds to a confined phase.

\subsection{Other Insulator states}\label{secother}

\begin{figure*}
    \includegraphics[width=18cm,height=4cm]{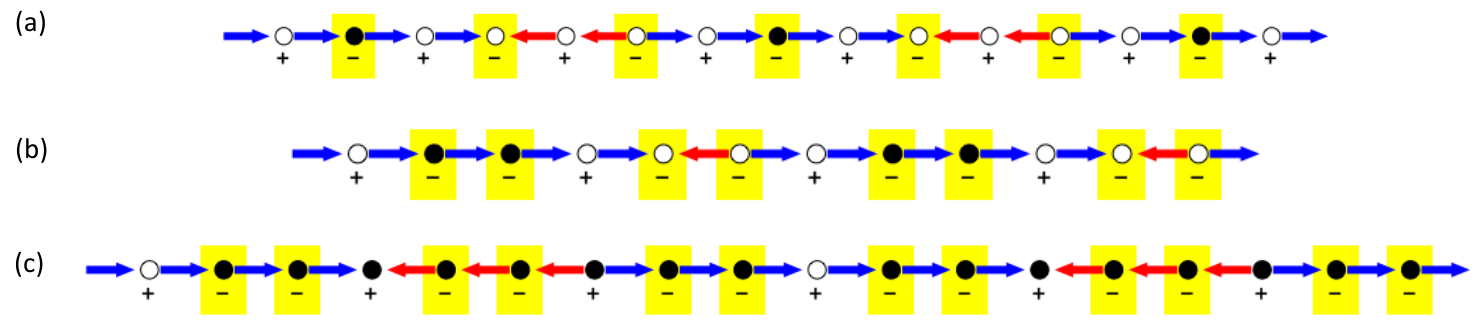}
    \caption{Configurations illustrating  (a) the 1/6th filling insulator state in the zero net magnetization sector $(+-+-+-)$,  (b) the  1/3rd filling insulator state and (c) the 8/9th filling insulator state in the finite net magnetization sector $(+--+--)$.  For large $h$ the bosons preferentially sit on the sites with $G_i=-1$.  The text gives an argument for the stability of these configurations.}
    \label{insulator}
\end{figure*}

In addition to the density waves in Sec.~\ref{sec:peierls},
we find incompressible states at other rational fillings $\nu=n_b/n_m$ shown in Fig.~\ref{fig:gap}. In particular, in the zero net magnetization Gauss law sector ($+-+-+-$), we find a finite charge gap at  $n_b=1/6$ and $5/6$, corresponding to $\nu=1/3$ and $5/3$.  
In the finite magnetization sector ($+--+--$), we find incompressible states at $n_b=1/3$ and $n_b=8/9$ corresponding to $\nu = 1/2$ and $\nu=4/3$. Cartoons of these states are shown in Fig.~\ref{insulator}.  
At large $h$ we can derive an effective model for these insulators.  Essentially the particles sit on a subset of the $-$ sites, forming a density wave, whose structure is set by an effective interaction between the particles.

For concreteness we focus on the $G=+--+--$ sector,  first analyzing the $n_b=1/3$ state.  Here we have two bosons for every six sites. Bosons preferably pair up on neighboring $'-'$ sites to minimize the Zeeman energy of the gauge spins.  In the absence of hopping, all such configurations are degenerate. However higher order hopping terms in $t/h$ break the degeneracy, leading to an incompressible state where every alternate pair of $'-'$ sites are occupied, as shown in the cartoon in Fig.~\ref{insulator}(b).

\begin{figure*}
    \includegraphics[width=15cm,height=7cm]{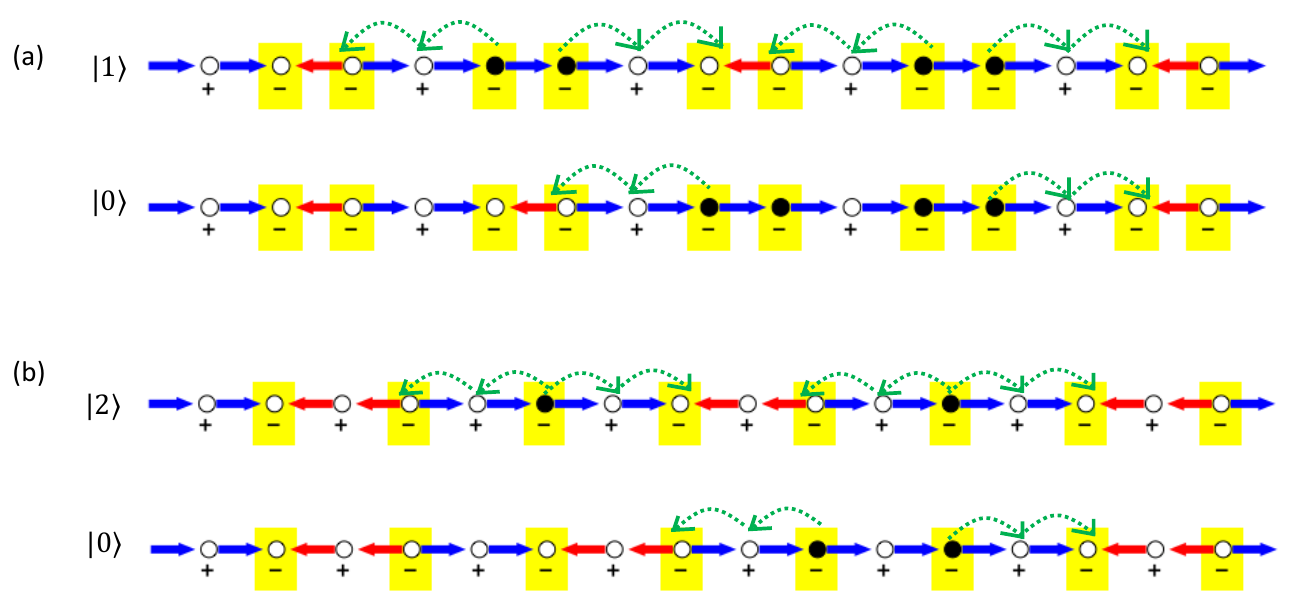}
    \caption{(a) Illustrative states, denoted $|1\rangle$ and $|0\rangle$, with one and zero pairs of unoccupied $'-'$ sites between the boson pairs at $-$ sites in the $G = +--+--$ sector. (b) Illustrative states, denoted $|2\rangle$ and $|0\rangle$, with two and zero unoccupied $'-'$ sites between the bosons at $-$ sites in the $G = +-+-+-$ sector. The curved green arrows show virtual hopping processes that break the degeneracy among the two pairs of configurations.}
    \label{fig:hop}
\end{figure*}

Fig.~\ref{fig:hop}(a) shows two low energy configurations where configuration $|1\rangle$ has one empty pair of $'-'$ sites between boson pairs while configuaration $|0\rangle$ has no empty pair of $'-'$ sites between bosons. These two configurations are connected by a sixth order hopping process with an effective hopping matrix element, $t_{\rm eff} = t^6/(128h^5)$, as $h\to\infty$. 
Virtual hopping processes, denoted by the green arrows in 
Fig.~\ref{fig:hop}(a), shift the relative energy of these two states.  There are more such processes for the  $|1\rangle$ state than there are for the $|0\rangle$ state, and therefore $E_1<E_0$.  
This virtual hopping is a fourth order process and perturbation theory yields the energy difference between $|0\rangle$ and $|1\rangle$ to be, $V_{\rm eff} =t^4/(8h^3)$. We can interpret this energy shift as interaction between nearest-neighbor boson pairs. We can therefore map this model to an XXZ model with hopping ($t_{\rm eff}$) and nearest neighbor-repulsion ($V_{\rm eff}$) of boson pairs. Since $V_{\rm eff} > 2t_{\rm eff}$, the ground state is an insulator~\cite{Giamarchi}.

We can do a similar analysis of the $n_b=1/6$ state in the $G_i = (-1)^{i}$ sector. Bosons preferably occupy only $'-'$ sites to minimize the Zeeman energy of the gauge spins. At this density, there is 
one boson for every three $'-'$ sites.  The spin-energy is minimized when particles  have  an even number of $-$ sites in between them.  Particle hopping breaks the degeneracy of all such low energy configurations and stabilizes the insulator state depicted in the cartoon picture in Fig.~\ref{insulator}(a).




Fig.~\ref{fig:hop}(b) shows two of the low energy configurations:  In configuration $|2\rangle$, two particles have two `$-$' sites (and three `$+$' sites) between them, while in configuration $|0\rangle$, they are on neighboring `$-$' sites, with only a single `$+$'
site between them. Sixth order perturbation theory yields an effective hopping matrix element, $t_{\rm eff} = t^4/(16h^3) -  t^6/(64h^5)$.  As in the prior argument, the configurations experience a relative energy shift coming from virtual processes in which a particle hops to other sites and then returns to its starting point.  As illustrated by the green arrows on Fig.~\ref{fig:hop}(b), there are fewer accessible virtual states when the particles are on neighboring `$-$' sites.  This leads to an energy separation between these states, which at sixth  order in perturbation theory is $V=E_0-E_2=2t^4/(16h^3)+2t^6/(64h^5)$.  Mapping this model to the XXZ model and given that $V_{\rm eff}>2 t_{\rm eff}$, the ground state is insulating \cite{Giamarchi}. 

This argument is readily generalized to sectors of the form $-++-++$, $-+++-+++$, and so on.  These will all have gapped insulators when $\nu=n_b/n_m=1/3$.  For the $-++-++$ sector, this corresponds to $n_b=1/9$.

It is worth emphasizing that our first example, the $n_b=1/3$ state in the $+--+--$ sector, has a much larger gap than our second example, the $n_b=1/6$ state in the $+-+-+-$ sector.  Both map onto a XXZ model, but the former has a much larger ratio of $U_{\rm eff}/t_{\rm eff}$.

We can produce further insights by using 
 the particle-hole transformation given in Sec.~\ref{model}.
 For example, the $n_b=1/6$ state in the $+-+-+-$ sector is transformed into the $n_b=5/6$ state in the same sector.  Thus our argument also explains the $n_b=5/6$ insulator seen in Fig.~\ref{fig:gap} and its associated charge density wave in Fig.~\ref{fig:fourier}.  In another zero net magnetization sector $G=-++-++$, there is an insulator at $n_b=1/9$ filling corresponding to $\nu = 1/3$. This is analogous to the $\nu =1/3$ insulator in the $G = +-+-+-$ sector. Under a particle-hole transformation, this yields an $n_b=8/9$ insulator in the finite net magnetization sector, $G=+--+--$.  That insulator is also seen in Fig.~\ref{fig:gap} (charge gap) and Fig.~\ref{fig:fourier} (charge density wave modulation).



\subsection{Single particle correlations}\label{sec:correlations}

\begin{figure}
    \includegraphics[width=\columnwidth]{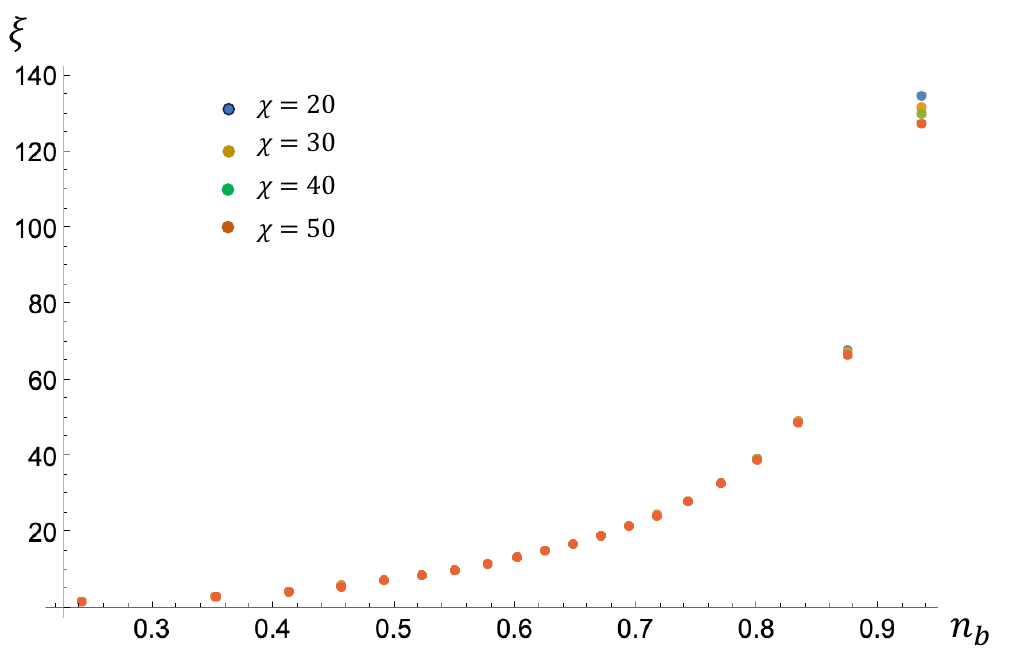}
    \caption{Single particle off-diagonal correlations, $C_{ij} = \langle a^{\dagger}_ia_j \rangle$ are fit to an exponential function ($C_{ij}=e^{-|i-j|/\xi}$). The plot shows dimensionless correlation length (measured as number of sites) $\xi$ as a function of boson density, $n_b$ for the uniform Gauss law sector, $G= ++++$ for bond dimension, $\chi =20,30,40,50$ using the VUMPS infinite DMRG algorithm. Correlation length $\xi$ appears to diverge only as $n_b \to 1$}
    \label{fig:uniformgauge}
\end{figure}


To characterize the compressible phases, we calculate the single-particle density matrix $\langle a_i^\dagger a_j\rangle$.  Generically we expect the bosons to be confined, in which case the correlations fall off exponentially at large distance, $\langle a^{\dag}_i a_j \rangle = a_0 e^{-|i-j|/\xi}$. Here $\xi$ is a dimensionless correlation length measured as number of sites. To calculate this correlation function, we make an infinite matrix product state ansatz for the ground state wavefunction, and use the VUMPS algorithm~\cite{vumps} in the Itensor library to optimize the parameters.  By directly working in the thermodynamic limit we avoid any boundary effects, but the bond dimension $\chi$ of the ansatz introduces an effective cutoff \cite{PhysRevB.105.134502}.  Thus we repeat the calculation with multiple $\chi$, and study how $\xi$ evolves.  For this calculation, we use the Gauss's law to eliminate the bosons and get a local Hamiltonian of just the gauge spins on the links. We don't explicitly impose  boson number conservation and instead use a chemical potential term to obtain an average boson density in the ground state (see Appendix~\ref{spinmodel} for calculation details).  Thus for each value of $\mu$ we calculate both $\xi$ and $n_b$.

Figure~\ref{fig:uniformgauge} shows the behavior of the single particle correlation length in the uniform sector, $G=+++++$.  As can be seen, for any density $n_b\neq 1$, the correlation length is finite, indicating that the particles are confined.  The correlation function diverges as $n_b\to 1$, indicating that the holes are deconfined for infinitesimal hole density.  In the graph we show the results from several different bond dimensions, illustrating convergence for all but the highest densities.  

These observations are consistent with the arguments in Sec.~\ref{sec:results}.  The uniform gauge has a net magnetization, and hence the energy is minimized if particles pair up spatially.  Near unity filling one can perform a particle-hole transformation, and map the system onto a small density of particles in the $G=-----$ gauge.  This has a net zero magnetization, so a small number of isolated particles are deconfined.  However, the numerics show that for any non-vanishing density they will be confined. 

We find similar results in other sectors:  For generic $n_b\neq 0,1$, we always find a finite correlation length.  The correlation length will diverge as $n_b\to 0$ if we are in a net zero magnetization sector.  Similarly, it diverges as $n_b\to 1$  if the net magnetization vanishes under a particle-hole transformation.  For example in the alternating section $G=+-+-+-$, the correlation length diverges both as $n_b\to 0$ and $n_b\to 1$.  We never see a diverging correlation length at the charge-density wave transitions.  This latter observation is sensible, as the effective model for excess particles/holes on top of the charge density waves always has a net magnetization.

\subsection{Supersolidity}

For generic filling we have a compressible state with no off-diagonal long range order in the single-particle density matrix.  Analogous to what has been seen in previous works, the two-particle density matrix, however, has long-range (power law) order, and the system is a pair superfluid~\cite{z2dimers}.  Additionally it is energetically favorable for particles to sit on sites where $G_i=-1$ to minimize the Zeeman energy of the gauge field spins.  Thus $O_{cdw}$ defined in Sec.~\ref{sec:peierls} is non-zero. Simultaneously having pair superfluidity and charge-density-wave order could be described as {\em supersolidity}.  One caveat is that the periodicity is imposed by the conserved quantities.  Nonetheless the Hamiltonian is translationally invariant.

\section{Summary and Outlook}
Experiments have begun to explore simple lattice gauge theories.  One important feature is that by changing their initial conditions they can access a range of different `Gauss law sectors', which have patterns of stationary charges.  Prior work has almost exclusively focused on the uniform sector.  We 
study ground state phases in periodic Gauss law sectors.  We classify them by the net magnetization of the gauge fields in the ground state in the absence of matter particles.  A small (infinitesimal) number of particles added to a finite magnetization sector will bind up into pairs, while in a zero magnetization sector they remain unpaired. These can be interpreted as confined and deconfined phases.  One can apply a particle-hole transformation and use equivalent reasoning to study a small (infinitesimal) number of holes -- finding both confined and deconfined states.
By numerically studying the single particle density matrix we find that as long as the density is not equal to $0$ or $1$ the particle/holes always pair and are confined, regardless of sector.   

At some special matter densities, we find  correlated insulators whose periodicity is either 
equal to that imposed by the static charges, or is a rational multiple.  We refer to the former as commensurate, and the latter as incommensurate.  The commensurate insulators are characterized by the filling factor $\nu=n_b/n_m=1$ where $n_b$ is the density of particles and $n_m$ is the density of static charges.  
We argue that the transition to the commensurate insulator is
a Peierls instability, and it occurs at infinitesimal $h$. The incommensurate density waves arise from the effective interaction between particles, in the presence of the periodic array of static charges.  

At generic filling we have a pair superfluid displaying a periodic density, despite the fact that the Hamiltonian is translationally invariant.  This can be considered a supersolid.


All of these features are accessible in current experiments.  The density waves are readily detected by imaging the density patterns.  Compressibility can be probed by looking at the density in a confining potential \cite{compressibility}.   Features of the single-particle density matrix can be explored through interference experiments \cite{interference} or time-of-flight expansion \cite{timeofflight}.  Confinement can be explored by observing the dynamics of isolated particles and pairs of particles \cite{exp2,sccircuit,exp4}.

Going forward, there are a number of other very interesting directions, particularly involving how dynamical quantities depend on the Gauss law sector.
Interesting questions include those involving
ergodicity and the presence of non-ergodic phenomena like quantum many-body scars~\cite{scars} in such periodic sectors.  Ultimately one would like to be able to explore phenomena related to high energy physics, such as particle production during various scattering events.

Any temporal gauge implementation of a  lattice gauge theory will have multiple Gauss law sectors.  This is neither limited to 1D nor to $Z_2$ lattice gauge theory.  
It would be exciting to study more complicated lattice gauge theories in non-uniform Gauss law sectors.

\section{Acknowledgement}
This material is based upon work supported by the National Science Foundation under Grant No. PHY- 2110250.

\appendix


 

 

\section{Inversion-Shift symmetric sectors}
\label{isss}
Although we did not explicitly make use of it in our analysis, it is useful to note that there are classes of gauge sectors for which one can replace the non-local many-body interaction in Eq.~(\ref{interaction}) with a local 2-body term plus a single-particle potential.

To define these classes, we first introduce
$\Lambda_i=\prod_{j\leq i} G_j$.  Simple examples include: $G=++++ \to \Lambda=+++++$, $G=-----\to \Lambda=+-+-+-$, and $G=+-+-+-\to \Lambda=++--++--++--$. 
In terms of $\Lambda$, and the occupation numbers $n_j$, the spin texture is
\begin{equation}\label{sig}
\sigma^x_{i,i+1} = (-1)^{\sum_{j\leq i} n_j} \Lambda_i 
\end{equation}
Our mapping onto a local potential works when there is some $s$ such that
$\Lambda_{i+s}=-\Lambda_i$.  For example $\Lambda=+-+-+-$ has this property with $s=1$ or $\Lambda=++--++--++--$ has $s=2$.

\begin{figure}
    \includegraphics[width=\columnwidth]{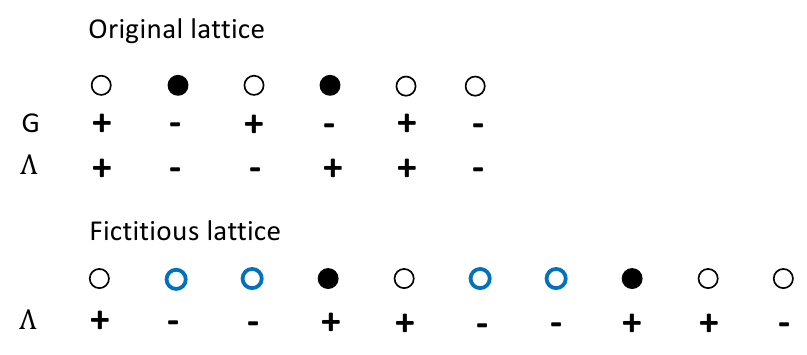}
    \caption{Upper panel shows the original lattice in the sector $G = +-+-+-$ showing sites occupied by bosons(dark). Lower panel shows the fictitious lattice with two auxilliary sites (blue) to the left of each occupied site.}
    \label{fig:fict}
\end{figure}

As illustrated in Fig.~\ref{fig:fict}, we define a fictitious lattice, where to the immediate left of each particle we add $s$ auxilliary sites, shown in blue in the figure.  That is, site $i$ on the real lattice corresponds to site $\bar i=i+  s \sum_{j\leq i} n_j$.  The configurations of particles on the original lattice are in 1-to-1 correspondence to the configurations on the new lattice where no particles are within $s$ sites of one-another.  In the fictitious lattice we therefore add a hard-rod constraint, which can be implemented via a two-body potential
\begin{equation}
U =\sum_{\bar i \bar j} W_{\bar i-\bar j} n_{\bar i} n_{\bar j}
\end{equation}
where $W_{\bar k}= \infty$ if $|\bar k|\leq s$, but otherwise $W_{\bar k}=0$.  Finally, the magnetic energy, can be written in terms of the degrees of freedom on the fictitious lattice by substituting Eq.~(\ref{sig}) into
$V=-h\sum_i \sigma^x_{i,i+1}$ and noting that $(-1)^m \Lambda_i=\Lambda_{i+s m}$.  Using that, we write
\begin{equation}
V=-h\sum_i \Lambda_{i+s \sum_{j\leq i} n_j}
\end{equation}
The sum over physical lattice sites $i$ can be replaced by a sum over the fictitious sites $\bar i$ which are not auxilliary sites.  The function $F_{\bar i}= 1-\sum_{t=1}^s n_{\bar i+t}$ has the property that it is zero on an auxilliary site, and is one on the original sites.  Hence
\begin{eqnarray}
V&=&-h\sum_{\bar i} \Lambda_{\bar i} F_{\bar i}\\
&=& - h \sum_{\bar i} \Lambda_{\bar i}
+ h \sum_{\bar i} \Lambda_{\bar i} \sum_{t=1}^s n_{\bar i+t}.
\end{eqnarray}
The first term is a constant, and the second can be written as $\sum_{\bar i} n_{\bar i} V_{\bar i}$ with
\begin{equation}
V_{\bar i} = h \sum_{t=1}^s \Lambda_{\bar i-t}.
\end{equation}
Thus, up to a constant, the Hamiltonian for the particles on the fictitious lattice is
\begin{equation}
H=\sum_{\bar j} -t(a_{{\bar j}+1}^\dagger a_{\bar j} + a_{{\bar j}}^\dagger a_{\bar j+1}) + V_{\bar j} n_{\bar j} +\sum_{\bar i \bar j} W_{\bar i-\bar j} n_{\bar i} n_{\bar j}.
\end{equation}

For example, if we take the $G=----$ sector, where $\Lambda=+-+-+-$, then $s=1$.  The potential $V=+h,-h,+h,-h$ is simply a sublattice bias in the fictitious lattice.  
There is also an infinite nearest neighbor repulsion from $W$. 
Due to this hard-rod constraint, the density in the fictitious lattice can be no greater than $1/2$.  One has a superfluid for any filling other than $0$ or $1/2$. 

As a less trivial example, consider $G=+-+-+-$ for which $\Lambda=++--++--++$ and $s=2$.  The potential in the fictitious lattice has period 4, $V=+2 h, 0, -2h, 0, +2h,0\cdots$, and there is an infinite nearest and next-nearest neighbor repulsion.  At large $h$, the $\nu=1$ state in Sec.~\ref{sec:peierls} corresponds to putting particles on each of the sites where $V=-2h$.  The $\nu=1/3$ state in Sec.~\ref{secother} corresponds to having one particle for every 8 sites of the fictitious lattice, ie. 1 particle for every 2 unit cells.  The $\nu=5/3$ state corresponds to 5 particle for every 16 fictitious sites (4 unit cells).  It is not completely obvious from this picture that the latter two configurations should be insulators, but the perturbative analysis from that section can readily be repeated in this language.

It is natural to ask how confinement is described in this fictitious lattice.  For simplicity consider the case $G=-----$, which we have already introduced.  In the fictitious lattice there is a period 2 superlattice, of amplitude $h$.  Clearly hard rod particles in such a potential are not confined.
Nonetheless, we know that the particles in the original lattice are confined.  These statements are not contradictory, as the single-particle density matrix of the particles on the physical lattice is not simply related to the single-particle density matrix of the particles on the fictitious lattice.  It is possible to have one of these functions fall off exponentially, while the other has algebraic correlations.

To understand this feature, consider the large $h$ limit, where we
will only have the alternate low energy sites of the fictitious lattice occupied.  In this limit there will always be an odd number of empty fictitious sites between any two neighboring particles.  In the original lattice this corresponds to an even number of holes between them.  Thus pairing of holes in the original lattice is manifest as occupation of the low energy sublattice in the fictitious lattice. 

It is useful to note that the $G=----$ sector maps onto the standard $G=++++$ sector under a particle-hole transformation.  Thus the uniform sector can be explored with this mapping.

\section{Spin model for calculating correlation functions}\label{spinmodel}

Here we present a variant of the argument from \cite{z2interactions}, which allows us to write the Hamiltonian in Eq.~\ref{hamiltonian} solely in terms of the spin degrees of freedom.  The key idea is that if we know the spin configuration in the $X$-basis ($\sigma^x_{i,i+1}$), and the values of $G_i$, we can construct the locations of all of the particles.  
This spin representation is convenient for numerical calculations.


We will formulate this transformation as a mapping between operators that act on the spin-particle space ($\sigma^{x}_{i,i+1},n_i$) and a spin chain ($\tau^{x}_i$).
For notational simplicity, in the pure spin model, we place the spins on sites, rather than bonds.  In the $x$-basis, the spin wavefunction is found by simply ignoring the particles, and hence $\tau_i^x=\sigma_{i,i+1}^x$.  Flipping a single spin changes the boson parity on the adjacent sites, and hence $\tau_i^z=(b_i+b_i^\dagger)\sigma_{i,i+1}^z(b_{i+1}+b_{i+1}^\dagger)$. Note that this term includes both a nearest-neighbor hopping and a pair creation/annihilation of bosons. Flipping a string of spins corresponds to adding or removing a single particle
$\prod_{j\geq i} \tau_j^z = (b_i+b_i^\dagger)\prod_{j\geq i} \sigma^z_{j,j+1}  = a_i+a_i^\dagger$, where the operators $a_i$ and $a_i^\dagger$ were introduced in Sec.~\ref{model} when we eliminated the spins.  They correspond to the simplest gauge invariant ways of adding or removing particles.  The hard core boson's annihilation operator $a_i$ vanishes on a state where no particles are on site $i$.  Thus
$a_i=(a_i+a_i^\dagger) (1-(-1)^{n_i})/2$.  We can then use Gauss' law, $G_i = \sigma^x_{i-1,i}(-1)^{n_i}\sigma^x_{i,i+1}$, to conclude

\begin{equation}
    a_i = (a_i+a_i^\dagger)(1-\sigma^x_{i-1,i}G_i \sigma^x_{i,i+1})/2
\end{equation}

\begin{equation}
    = [\prod_{j\geq i} \tau_j^z ][(1-\tau^x_{i-1}G_i \tau^x_{i})/2]
\end{equation}

\begin{equation}
    = [(1+\tau^x_{i-1}G_i \tau^x_{i})/2]
[\prod_{j\geq i} \tau_j^z ]
\end{equation}


where the last line comes from commuting the chain of $\tau^z$'s past the $\tau^x$'s.
Similarly, 

\begin{equation}
    a_i^\dagger = [ (1-\tau^x_{i-1}G_i \tau^x_{i})/2 ][\prod_{j\geq i} \tau_j^z]
\end{equation}

\begin{equation}
=[\prod_{j\geq i} \tau_j^z][ (1+\tau^x_{i-1}G_i \tau^x_{i})/2 ].
\end{equation}

The hopping term in the Hamiltonian,
$K_{i,i+1}=-(b_{i+1}^\dagger \sigma^z_{i,i+1}b_i + b_{i}^\dagger \sigma^z_{i,i+1}b_{i+1})=-(a_{i+1}^\dagger a_i + a_i^\dagger a_{i+1})$ 
can then be written as
$K^s_{i,i+1}=(G_i G_{i+1}\tau^x_{i-1} \tau^x_{i+1} - 1) \tau^z_{i}/2$.


 One complication with this spin representation is that number conservation is somewhat less transparent.  Thus it is convenient to work in the grand Canonical ensemble, introducting a chemical potential, $\mu$.  The number of particles is
 \begin{equation}
\hat N= \sum_i (1-G_i \tau_{i-1}^x \tau_{i}^x)/2.
 \end{equation}
 Up to an irrelevant constant, the resulting grand canonical Hamiltonian is
 \begin{equation}\label{soh}
 \begin{split}
\hat H-\mu \hat N
=-\frac{t}{2}\sum_i \left( \tau^x_{i-1} G_{i} G_{i+1}\tau^x_{i+1}-1\right) \tau^z_{i}\\-h\sum_i \tau_{i}^x
+\frac{\mu}{2}\sum_i  \tau^x_{i-1}G_i\tau^x_{i}
\end{split}
 \end{equation}
This spin-only Hamiltonian consists solely of local terms, in contrast to the highly non-local many-body interaction term we found in Eq.~\ref{interaction} after eliminating gauge spins. For the uniform Gauss law sectors, where all $G_{i}$ are either $+1$ or $-1$, this model maps onto the 
cluster-state Hamiltonian~\cite{clusterstate}.


Since $\hat N$ commutes with $\hat H$, the ground state of $\hat H-\mu \hat N$ has a fixed density of particles, controlled by $\mu$.
When $\mu$ is large and negative the grounds state has $n=0$ on each site, while when $\mu$ is large and positive it has $n=1$.


In our infinite matrix product state calculation using the VUMPS algorithm, we find the ground state of Eq.~\ref{soh}. We tune the value of $\mu$, and calculate both $n$ and the correlation length,  as displayed in Fig.~\ref{fig:uniformgauge}.
The single particle correlation function, can be formulated in terms of gauge spins as
\begin{equation}
\langle a_i^{\dagger}a_j\rangle=
\left\langle 
\left(\frac{1-\tau^x_{i-1}G_i\tau^x_i}{2}\right)
\left(
\frac{1+\tau^x_{j-1}G_j\tau^x_j}{2}\right)
\prod_{i<k<j} \tau^z_{k}\right\rangle.
\end{equation}


\bibliography{lgt.bib}

\end{document}